\documentclass[10pt,aps,pra,reprint,superscriptaddress,longbibliography]{revtex4-1}
\usepackage{header}

\begin{document}

\title{Quantum Hamiltonian Evolution for Coherent Quantum Learning}

\author{Ignacio B. Acedo\,\orcidlink{0009-0006-3860-6923}}
\email{ibenito@quantum-mads.com} 
\affiliation{Quantum Mads, Calle Larrauri 1, Edificio A, piso 3, puerta 28, 48160 Derio, Spain}

\author{Javier Gonzalez-Conde\,\orcidlink{0000-0003-3385-7943}}
\affiliation{Quantum Mads, Calle Larrauri 1, Edificio A, piso 3, puerta 28, 48160 Derio, Spain}
\affiliation{Department of Physical Chemistry, University of the Basque Country UPV/EHU, Apartado 644, 48080 Bilbao, Spain}
\affiliation{EHU Quantum Center, University of the Basque Country UPV/EHU, Apartado 644, 48080 Bilbao, Spain}

\author{Pablo Rodriguez-Grasa\,\orcidlink{0009-0009-8622-2088}}
\affiliation{Department of Physical Chemistry, University of the Basque Country UPV/EHU, Apartado 644, 48080 Bilbao, Spain}
\affiliation{TECNALIA, Basque Research and Technology Alliance (BRTA), 48160 Derio, Spain}

\author{Barry C. Sanders\,\orcidlink{0000-0002-8326-8912}}
\affiliation{\mbox{Institute for Quantum Science and Technology, University of Calgary, Alberta T3A 0E1, Canada}}

\author{Lirandë Pira\,\orcidlink{0000-0002-6305-1150}}
\email{lpira@nus.edu.sg}
\affiliation{Centre for Quantum Technologies, National University of Singapore, Singapore}

\date{\today}

\begin{abstract}
We introduce Coherent Quantum Learning (CQL), a training framework for quantum learning models in which the model parameters are quantum degrees of freedom evolved under a Hamiltonian that encodes the loss function. Current quantum machine learning retains classical optimization: parameters are updated by a classical outer loop using gradient estimates from measurements, and quantum coherence has no role in the training dynamics, just as in any classical treatment of the same problem. In the quantum case, a parameter register initialized in superposition evolves unitarily, and probability amplitude concentrates near low-loss configurations through interference, without gradient computation or classical feedback. We give an explicit construction using block encodings and Hamiltonian simulation, applicable to arbitrary parameterized circuits. Numerical experiments on binary classification and interferometric phase estimation confirm that the evolved distribution peaks at the optimal parameters, matching gradient-based performance. The construction is compatible in principle with fault-tolerant implementations and extends to batched training via sequential Hamiltonian evolution.
\end{abstract}

\maketitle
\twocolumngrid

\textit{Introduction.---}
Learning is the physical process by which a system acquires information from data. In machine learning, this process is typically formulated as optimization: model parameters are iteratively updated to minimize an objective function constructed from observations~\cite{bishop2006pattern,murphy2012machine,rumelhart1986learning,lecun2015deeplearning,goodfellow2016deep}. From a physical perspective, these updates define a dynamical evolution in parameter space, where the state of the learning system evolves toward increasingly informative representations. Quantum computing, in turn, describes information processing through coherent unitary evolution of quantum states~\cite{manin2007mathematics,benioff1980computerasaphysicalsystem,feynman1982simulatingphysics,NielsenChuang2000}. This naturally raises the question of whether learning itself can be formulated as a quantum physical process, rather than as a classical optimization algorithm applied to a quantum model.

Most existing approaches to quantum machine learning retain the classical paradigm for training~\cite{Biamonte2017,SchuldPetruccione2018}. Parameterized quantum circuits are evaluated at individual parameter settings, gradients are estimated from repeated measurements using techniques such as the parameter-shift rule~\cite{mitarai2018quantumcircuitlearning,schuld2019evaluating,wierichs2022generalparametershift}, and parameter updates are performed by a classical optimizer~\cite{schuld2020circuitcentric,Kubler2020adaptiveoptimizer,stokes2020quantumnatural, Sweke2020stochasticgradient,cerezo2021variationalquantumalgorithms}. Although the model itself is quantum, the learning dynamics remain entirely classical: each parameter configuration is explored independently, and quantum superposition and interference play no role in the training process.

There are both practical and fundamental reasons to reconsider this paradigm. Hybrid optimization inherits statistical uncertainty from quantum measurements, and expressive parameterized circuits may exhibit barren plateaus in which gradients vanish exponentially with system size~\cite{mcclean2018barrenplateaus,larocca2025barrenplateaus}. More fundamentally, recent work has shown that classical backpropagation, which underpins the training of most modern neural networks~\cite{rumelhart1986learning} cannot be directly transferred to quantum systems without access to multiple copies of the quantum state, because measurement collapse prevents the reuse of intermediate information~\cite{benedetti2019parameterizedquantumcircuits,quantum_backpropagation_1,Bowles2025backpropagation}. These observations have motivated alternative approaches, including fully quantum training protocols, coherent parameter updates and gradient-free quantum optimization schemes~\cite{beer2020training, verdon2018universaltrainingalgorithmquantum,ye2025quantumautomatedlearningprovable}, suggesting that quantum learning may require training principles that are native to quantum mechanics rather than inherited from classical optimization.

A particularly promising direction is to formulate optimization itself as quantum dynamics~\cite{leng2023quantumhamiltoniandescent,chen2025quantumlangevindynamics}. Building on variational formulations of accelerated optimization based on the Bregman--Lagrangian~\cite{variational_perspective} and their Hamiltonian interpretation~\cite{betancourt2018symplecticoptimization}, quantum Hamiltonian descent~\cite{leng2023quantumhamiltoniandescent,leng2025quantumhamiltoniandescentnonsmooth} replaces a single classical optimization trajectory by Schrödinger evolution in parameter space, where superposition, interference, and tunneling collectively drive the wavefunction toward low-loss regions. Rather than updating parameters through an external optimization routine, the optimization process is realized through Hamiltonian evolution, which can be implemented on a quantum computer using modern Hamiltonian simulation techniques~\cite{Low_2017,low2019hamiltonian,trotter,berry2015simulating,PRXQuantum.6.010359}. Beyond its conceptual appeal, recent theoretical results suggest that Hamiltonian-based optimization can provide provable computational advantages over classical optimization in certain settings~\cite{catli2025exponentiallybetterboundsquantum}.

In this Letter, we introduce \emph{Coherent Quantum Learning} (CQL), a general framework in which the training of parameterized quantum models is itself realized as coherent quantum evolution. Model parameters are promoted to quantum degrees of freedom and evolve under a Hamiltonian whose potential encodes the learning objective. As the parameter state evolves, interference naturally amplifies configurations associated with low-loss, eliminating the need for explicit gradient computation or classical feedback. Unlike general optimization frameworks, CQL is designed specifically for learning problems, where the objective function is generated by evaluating a parameterized quantum model on data. Consequently, the training dynamics become data dependent and learning itself is implemented through Hamiltonian evolution.

This perspective extends naturally beyond conventional machine learning. While CQL applies directly to the training of variational quantum models, the same framework also describes physical inference tasks in which unknown parameters are encoded in quantum dynamics. We demonstrate this versatility through two representative examples: coherent training of a quantum classifier and coherent learning of an unknown interferometric phase. These two examples suggest a broader viewpoint in which learning from data and learning physical parameters are both instances of the same underlying physical process.\\

\textit{Coherent Quantum Learning (CQL) protocol.---} We now introduce the protocol for coherent training a quantum learning model. Consider a parameterized quantum circuit with trainable parameter $\theta$ and a loss function $\mathcal{L}(x,\theta)$ evaluated on input data $x$. To enable coherent training, we discretize the parameter $\theta$ using $m$ bits, $
\theta = 2\pi \sum_{j=1}^{m} \theta_j 2^{-j}, \  \ \theta_j \in \{0,1\},$
which yields a grid $\mathcal{G}_m$ of $2^m$ possible values in $[0, 2\pi)$, each represented by a computational basis state $\ket{\theta} = \ket{\theta_1 \theta_2 \dots \theta_m}$ on $m$ ancillary qubits. Therefore, one can define the corresponding discrete position and momentum operators in the $\theta$ basis denoted as $\hat \theta,\ \hat P_\theta$ respectively.

For simplicity we will assume the batch is of size $1$, where the extension to a larger size is straightforward. We build upon the time evolution introduced in~\cite{leng2023quantumhamiltoniandescent},
\begin{equation}\label{eq:ham_CQL}
    \hat{H}_\text{CQL}(t) =e^{\varphi(t)}\left(-\frac{1}{2}\hat{P}^2_\theta\right) + e^{\chi(t)}\hat H_{\mathcal{L}}(x,\hat \theta),
\end{equation}
where $e^{\varphi(t)}$, $e^{\chi(t)}$ are the damping parameters, $\hat{P}^2_\theta$ is the Laplacian operator in the $\theta$ basis and $\hat H_{\mathcal{L}}(x,\hat \theta)$ is the diagonal Hamiltonian proportional to $\mathcal{L}(x,\theta)$ that encodes the objective function to minimize. In the absence of prior information, the initial condition is chosen as an equal superposition over the parameter space, ensuring an unbiased exploration of all configurations. If prior information is available (e.g., an informed initial guess $\theta_0$), it can be incorporated by initializing the state as a localized wavepacket (e.g., a Gaussian) centered around $\theta_0$, thereby biasing the evolution toward promising regions of the landscape.
\begin{figure}[t]
    \centering
    \includegraphics[width=.99\columnwidth]{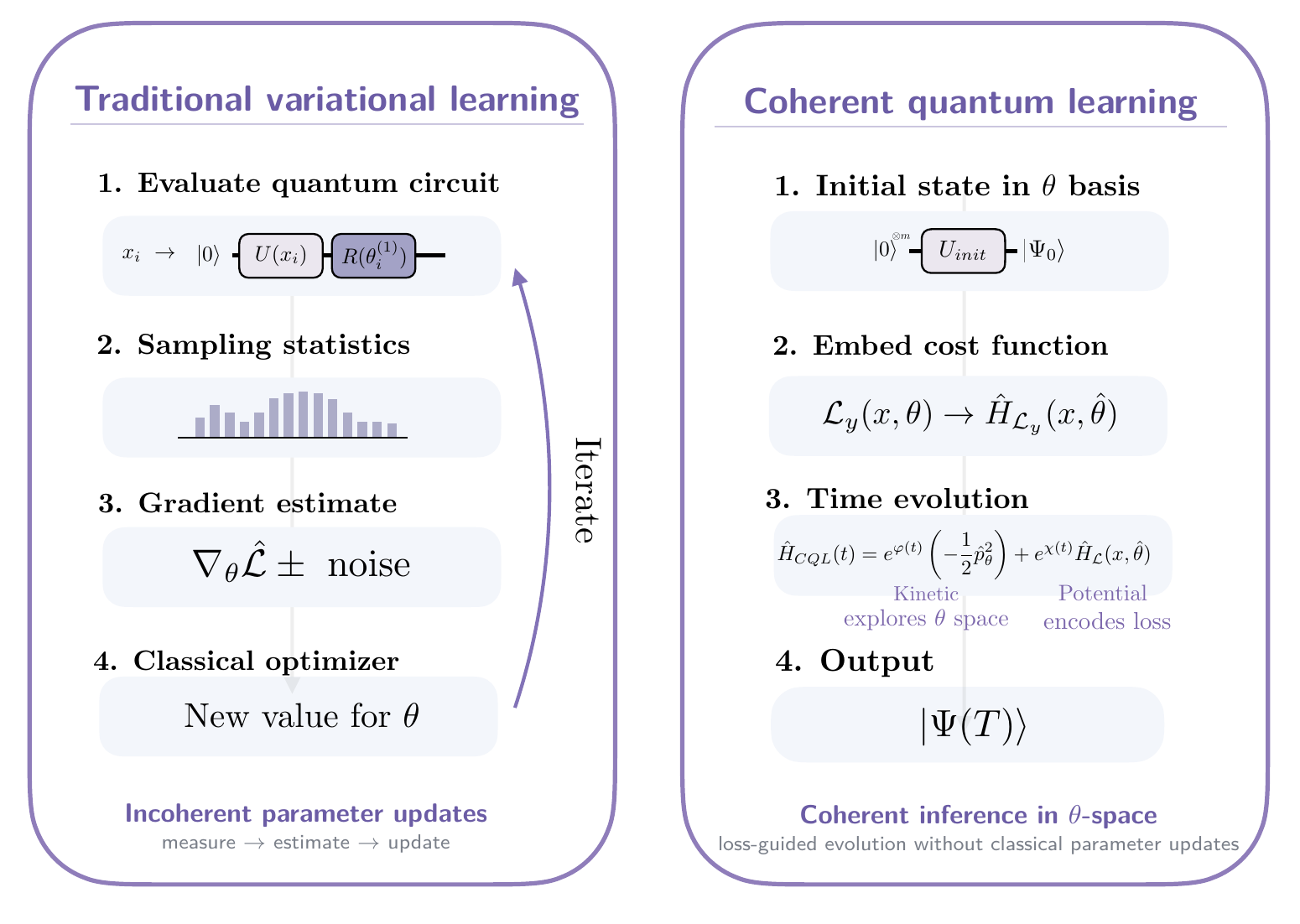}
    \caption{Conceptual comparison between hybrid training and Coherent Quantum Learning (CQL). In the traditional approach (left), a parameterized quantum model is evaluated at individual parameter settings, and measurement outcomes are used by a classical optimizer to iteratively update the parameters. In CQL (right), the model parameters are promoted to quantum degrees of freedom, the objective function is encoded as a loss Hamiltonian, and training is realized through coherent Hamiltonian evolution.}
    \label{fig:figure_1}
\end{figure}
The first obstacle to simulating the dynamics efficiently on a gate-based quantum computer arises from the time dependence of the Hamiltonian~\cite{wiebe2011simulating,wiebe2011quantumcomputer}. Simulating time-dependent Hamiltonians is not a native operation in the circuit model, and typically introduces considerable overhead~\cite{gonzalezconde2025costofemulating}. To address this issue, we employ the technique outlined in Refs.~\cite{cao2023quantum,cao2024unifying}, which involves enlarging the Hilbert space, introducing a ``clock dimension''~\cite{page1983page}. This expansion transforms the time dependence of a given Hamiltonian into a position operator $\hat{X}_t$ in an additional Hilbert space. Consequently, albeit at the expense of enlarging the Hilbert space to $\mathcal{H}_t\otimes\mathcal{H}_\theta$, this approach enables us to work with a time-independent Hamiltonian
\begin{equation}\label{eq:Hamil_time_indep}
\hat{H}_\text{CQL}=\hat{P}_t\otimes \mathbb{I}+e^{\varphi(\hat{X}_t)}\otimes\left(-\frac{1}{2}\hat{P}^2_\theta\right) + e^{\chi(\hat{X}_t)}\otimes \hat H_{\mathcal{L}}(x,\hat\theta),
\end{equation}
where we assume $\hat{X}_t,\ \hat{P}_t$ to be discretized by using $n_t$ ancilla registers and $\ket{t_0}$, the initial state for the new register, to be a representation of a localized wave function (delta function) in the position basis at $x_t=0$. In this way, the complexity of training our learning model is transferred to the simulation of the dynamics generated by Eq.~(\ref{eq:Hamil_time_indep}). Throughout this Letter, we assume access to a block encoding of the Hamiltonian, and explicitly describe how to construct the novel components.

\begin{figure}[t!]
     \centering
     \includegraphics[width=.9\columnwidth]{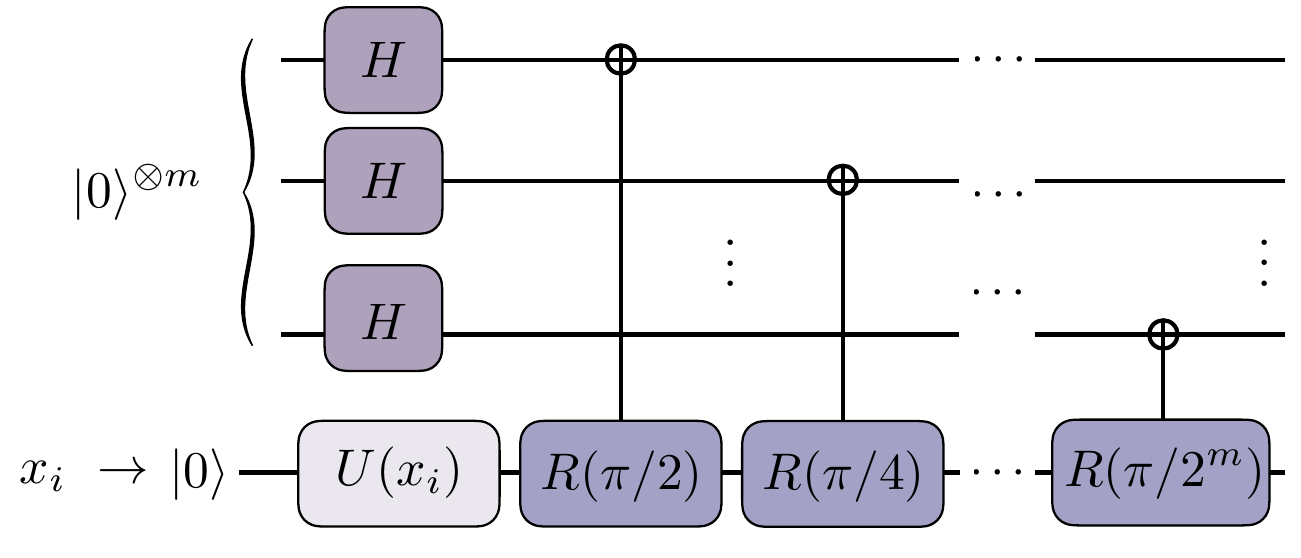}
     \caption{Circuit for one coherent training step. An ancilla register controls discretized parameter shifts and encodes the loss as phase.}
     \label{fig:figure_2}
\end{figure}

We firstly address how to construct the block encoding of $\hat H_{\mathcal{L}}(x,\hat \theta)=\sum_{\theta\in \mathcal{G}_m}\mathcal{L}(x,\theta)\ket{\theta}\bra{\theta}$. The key ingredient in this construction is a coherent model-evaluation unitary $U_{x,\theta}$ that evaluates the learning model over all parameter values of the grid simultaneously. To fix ideas, consider a reduced version of the data re-uploading model~\cite{perezsalinas2020datareuploading} for binary classification. Given an input $x \in \mathbb{R}$ with label $y \in \{0,1\}$, the model prepares the one-qubit state
\begin{equation}
\ket{\psi(x,\theta)} = R_y(\theta)\,U(x)\ket{0}
= \alpha_0(x,\theta)\ket{0} + \alpha_1(x,\theta)\ket{1},
\end{equation}
where $U(x)$ is a data-encoding unitary (e.g., $R_x(x)$) and $R_y(\theta)=e^{-i\theta Y/2}$ is a trainable rotation. The cost function to be minimized is defined as
\begin{equation} \label{eq:cost_function}
\mathcal{L}_y(x,\theta) = 1 - |\alpha_y(x,\theta)|^2,
\end{equation}
where $\alpha_y = (1-y)\alpha_0+ y\alpha_1$, which penalizes incorrect classification outcomes. Crucially, by adding the $m$ ancillary qubits encoding the parameter register, the model can be evaluated coherently over all parameter values of the grid by a unitary $U_{x,\theta}$ acting on the main qubit and the ancillas,
\begin{equation}\label{eq:superposition_ev_theta}
U_{x,\theta} \bigl(\ket{0}\otimes\ket{0}^{\otimes m}\bigr)
= \sum_{\theta \in \mathcal{G}_m} \Bigl[
\alpha_0(x,\theta)\ket{0}\ket{\theta}
+ \alpha_1(x,\theta)\ket{1}\ket{\theta}
\Bigr].
\end{equation}
This operation entangles model outputs with parameter configurations on a $m+1$ qubit state; the explicit circuit implementation is depicted in Fig.~\ref{fig:figure_2}. Note that we can evaluate simultaneously more than one data point by adding one ancilla per data, thus increasing the size of the batch. In this case, the amplitude that defines the cost function, Eq. (\ref{eq:cost_function}), is given by the bit-string of the corresponding labels, which can be mapped to the string $\ket{0 ... 0}$ by applying $X$ gates in those registers where the label is 1.

Without loss of generality we can assume $y=1$, which translates into $\mathcal{L}(x,\theta) = |\alpha_0(x,\theta)|^2$. Therefore, the next step is to achieve a block encoding of $\hat H_{\mathcal{L}}$ from query access to $U_{x,\theta}$. To this end, we can use Theorem 2 from Ref.~\cite{rattew2023nonlinear} in order to prepare a $(1, m+4, 0)$ block encoding $U_{\sqrt{H}}$ of the diagonal operator $\sqrt{H}=\sum_{\theta \in \mathcal{G}_m} \alpha_0(x, \theta) \ket{\theta} \bra{\theta}$ with $\mathcal{O}(m)$ circuit depth and $\mathcal{O}(1)$ queries to a controlled version of $U_{x,\theta}$. Note that the ancilla used for classifying $x$ has been counted as an additional ancilla qubit in the block encoding. By using the same protocol but this time applied to $U^\dagger_{x,\theta}$ instead, we get $U_{\sqrt{H}^\dagger}$, a $(1, m+4, 0)$ block encoding of $\sqrt{H}^\dagger=\sum_{\theta \in \mathcal{G}_m} \alpha^*_0(x, \theta) \ket{\theta} \bra{\theta}$. Using these two block encodings, we now build the block encoding of $\hat H_{\mathcal{L}}$, which actually can be written as a block encoding of $\sqrt{H}\sqrt{H}^\dagger$. To achieve the block encoding of a product from the individual block encodings of same the size~\cite{dalzell2025quantumalgorithms}, we can use a single ancillary qubit, see Supplemental Material. We thereby obtain a $(1, m+5, 0)$ block encoding of $\hat H_{\mathcal{L}}$ with $\mathcal{O}(m)$ circuit depth and $\mathcal{O}(1)$ queries to a controlled version of $U_{x,\theta}$.

Further, in order to achieve the block encoding of $\hat{H}_\text{CQL}$, we need to obtain the block encoding of certain functions of $\hat x_t,\ \hat{P}_t,\ \hat{P}_\theta$. The discrete representation of these operators, broadly studied, is not unique, as it depends on the discretization, the numerical method employed, and the boundary conditions. Some explicit examples on how to construct the block encoding of these operators can be found in Ref.~\cite{guseynov2025gateconstructionofblockencodings}. A systematic analysis of optimal discretizations and the control of operator norms to avoid unfavorable scaling is beyond the scope of this Letter.

Next, one may apply the Quantum Singular Value Transformation (QSVT)~\cite{gilyen2019quantumsingularvaluetransformation} to realize a polynomial transformation to $\hat x_t$ approximating the functions $e^{\varphi(\hat{X}_t)},\ e^{\chi(\hat{X}_t)}$, obtaining the corresponding block encodings. Alternatively, we can consider standard techniques for block-encoding arbitrary diagonal operators, e.g., via quantum read-only memory for classical function values of $\chi$ on the $\hat{X}_t$ register, or via the amplitude-to-diagonal conversion of Theorem 2 of Ref.~\cite{rattew2023nonlinear} when the function values can be loaded as state amplitudes. After homogenizing all block encodings in the same Hilbert space to act in the same number of ancillas, the target Hamiltonian can be constructed as a linear combination of the individual block encodings.

Finally, since access to our Hamiltonian is provided via a block encoding, we use qubitization~\cite{low2019hamiltonian} to implement the time evolution of the initial state. In this way, we obtain a quantum state that encodes a probability distribution on the values of $\theta$ whose maximum corresponds to the optimal value for training our learning model.

\begin{figure*}[t]
    \centering
    \includegraphics[width=\linewidth]{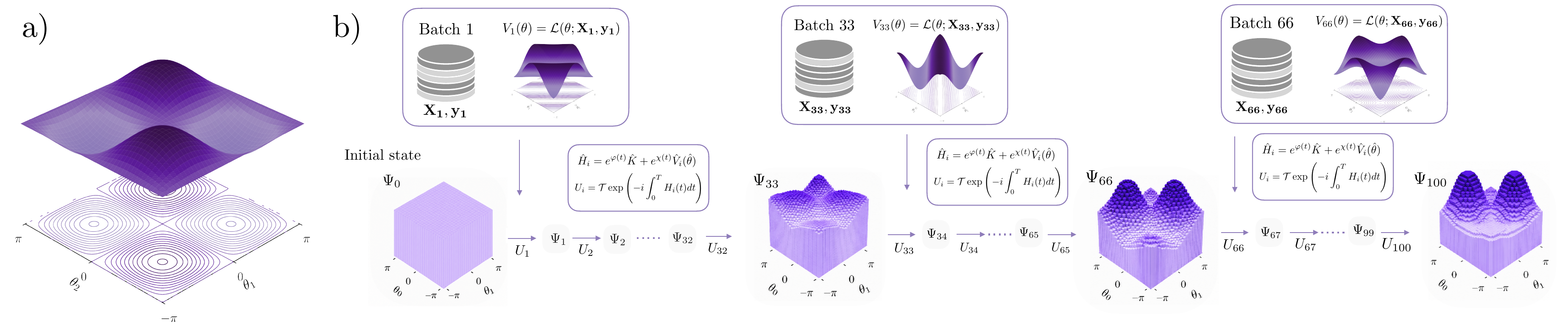}
    \caption{
    Numerical illustration of batch-based Coherent Quantum Learning (CQL). (a) Loss landscape $\mathcal{L}(\theta; \mathbf{X}, \mathbf{y})$ evaluated over the full training dataset as a function of the two variational parameters $\theta_0, \theta_1$, with contours indicating the structure of the global objective. (b) Evolution of the probability distribution of the parameter state $\hat{\theta}$ across sequential batched coherent training. Starting from a uniform superposition $\Psi_0$ over parameter space, the state is evolved under the CQL Hamiltonian, where the loss potential is constructed from individual training samples (batch size $1$). The loss landscape associated with select batches is shown alongside the corresponding parameter distributions, and the final state $\Psi_{100}$ after processing all $100$ batches. As training progresses, the probability amplitude increasingly concentrates in regions of low-loss, and after completing the full epoch, the final distribution anti-correlates with the global loss landscape shown in (a), confirming that the coherent evolution has effectively learned an inverse representation of the objective function.}
   \label{fig:batch_illustration}
\end{figure*}

\textit{Batched CQL.---}  So far, we have introduced general coherent training for a single batch. Although it is possible to construct a batch the size of the entire dataset, corresponding to coherent global training, this would require as many ancillas as data points. This is not a realistic scenario, as it becomes impractical for large datasets. Therefore, the need for iterative learning from different batches arises.  We propose an iterative batched training strategy, analogous to mini-batch methods in classical machine learning, whereby the dataset is partitioned into smaller subsets that are processed sequentially. 

Let us consider a training set divided into $B$ batches. Starting from an initial parameter state, the system evolves under the Hamiltonian in Eq.~(\ref{eq:ham_CQL}) using the first batch of training samples. The resulting parameter state is then used as the initial condition for a subsequent evolution under an updated Hamiltonian that encodes the next batch into the data registers. This sequential process is repeated until all batches in the training set have been processed. Here, the choice of batch size introduces a fundamental trade-off between spatial resources, namely the qubit count, and temporal resources, namely the circuit depth. Fig. ~\ref{fig:batch_illustration} shows a schematic representation of batched coherent training and is discussed in more detail later.

The main challenge of batch training is for the quantum system to retain information from previous batches while, at the same time, learning from new samples. This requirement fundamentally alters the role of the damping schedules compared to the quantum Hamiltonian descent framework~\cite{leng2023quantumhamiltoniandescent}. In the conventional global optimization setting, the schedules $\varphi(t)$ and $\chi(t)$ are typically chosen such that $e^{\varphi(t)/\chi(t)} \rightarrow 0$ as $t \rightarrow \infty$, thereby concentrating the parameter-state distribution around the global minima of the loss landscape. In the batched setting, however, such aggressive concentration is generally undesirable, since it would overemphasize the information contained in the current batch and reduce robustness to subsequent batches while erasing information from previous batches. 

This phenomenon is directly analogous to stochastic or mini-batch gradient methods, where the update magnitude is regulated through a learning rate hyperparameter. In the present framework, the analogous control parameter is the effective evolution time under the CQL Hamiltonian, together with the associated damping schedules. These quantities determine how strongly the parameter distribution adapts to each processed batch, and therefore govern the balance between local batch fitting and global generalization.  Consequently, in the batched regime, the asymptotic divergence condition on the ratio $\varphi(t)/\chi(t)$ is no longer appropriate and need not be imposed.

\textit{Numerical demonstrations.---}
(a) \textit{Quantum classification with a data re-uploading model}.
To illustrate the proposed protocol, we consider a binary classification task in which the objective is to determine whether input points lie inside or outside a one-dimensional interval. As a quantum model, we employ a single-qubit data re-uploading circuit with two layers of trainable rotations, yielding a variational ansatz with two parameters. The corresponding loss landscape is shown in Fig.~\ref{fig:batch_illustration}(a). Each parameter is then discretized using five qubits. Within this framework, we perform one training epoch using the coherent training protocol in the single-sample regime.

Fig.~\ref{fig:batch_illustration}(b) presents the evolution of the probability distribution induced by the discretized variational parameter space. Starting from an initial uniform superposition, the probability distribution progressively adapts to the structure of the cost function, developing pronounced peaks at parameter values associated with minima of the loss landscape. As training proceeds, the distribution increasingly reflects the geometry of the underlying objective function, ultimately exhibiting a bimodal concentration around the dominant minima. These results indicate that the model effectively learns an inverse representation of the loss landscape encoded in the final probability amplitudes. Rather than producing a single optimized parameter value, the protocol generates a quantum representation of the landscape itself, where low-loss regions are encoded as high-probability components of the final state. Further details regarding the exact configuration of these numerical results are provided in Appendix~\ref{sec:classification_details}.

The computational complexity of CQL depends on the cost of Hamiltonian simulation and the construction of the loss Hamiltonian. While a complete complexity analysis is beyond the scope of this work, the framework naturally inherits advances in block encoding and Hamiltonian simulation. Determining whether coherent learning can provide asymptotic advantages over hybrid optimization remains an important open question.

(b) \textit{Interferometric phase learning}.
To demonstrate the applicability of CQL to physical systems and parameter estimation, we ground our framework in a concrete physics-motivated task: learning an unknown phase shift within a two-mode interferometer (such as a Mach--Zehnder setup). Unlike the classification example, this task does not arise from machine learning but from a physical inference problem. It illustrates that the same coherent learning protocol naturally applies to learning unknown Hamiltonian parameters. 

\begin{figure}[t]
    \centering
    \includegraphics[width=\linewidth]{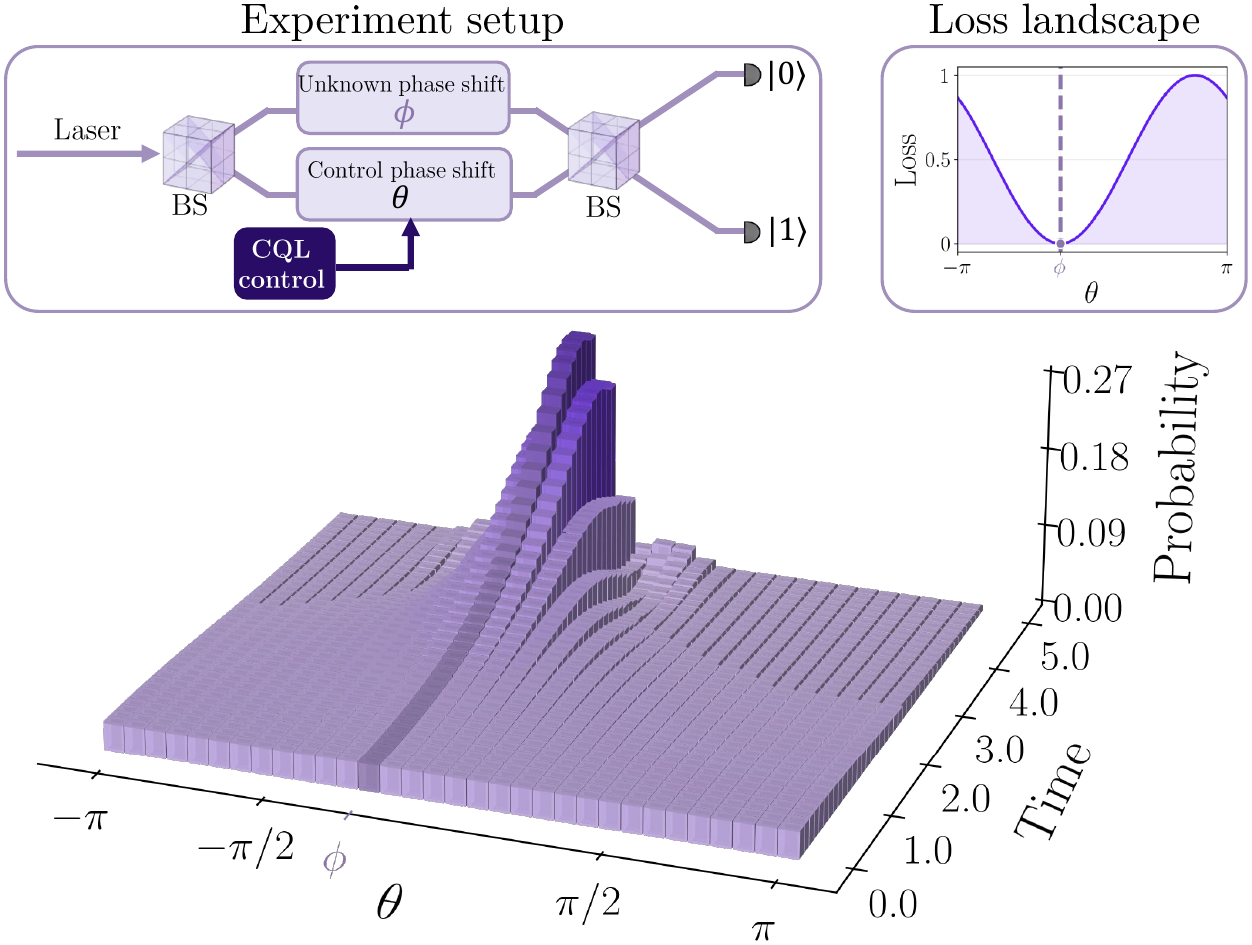}
    \caption{
    Upper left: Schematic of the interferometric phase-estimation setup. A single photon propagates through a balanced Mach--Zehnder interferometer containing an unknown phase shift $\phi$ and a controllable phase shift $\theta$, which is controlled by the CQL protocol. Upper right: Corresponding loss landscape, $\mathcal{L}(\theta,\phi)=\sin^2[(\theta-\phi)/2]$, whose global minimum identifies the unknown phase. Main: Time evolution of the probability distribution over the discretized parameter register, $|\Psi(\theta,t)|^2$. Starting from a uniform superposition, the coherent dynamics progressively localize the probability amplitude around the global minimum of the loss landscape, yielding an accurate estimate of the unknown phase $\phi$.
    }
    \label{fig:interferometer_results}
\end{figure}

The physical setup, illustrated in the upper-left panel of Fig.~\ref{fig:interferometer_results}, consists of a balanced Mach--Zehnder interferometer~\cite{Zehnder1891, Mach1892}. A single photon is split by a $50:50$ beam splitter (BS) into two spatial modes, defining the computational basis states $|0\rangle$ (lower arm) and $|1\rangle$ (upper arm). The upper arm contains an unknown phase shift $\phi\in[-\pi,\pi)$, while the lower arm contains a controllable phase shift $\theta$, which is dynamically updated by the CQL protocol. The two paths are subsequently recombined at a second balanced beam splitter before projective measurement at the output ports.

The interference pattern depends solely on the relative phase between the two arms, yielding the measurement probabilities $p(0) = \sin^2\left(\frac{\phi - \theta}{2}\right)$ and $p(1) = \cos^2\left(\frac{\phi - \theta}{2}\right)$ (see Appendix~\ref{sec:interferometer_details} for the full algebraic derivation). Consequently, estimating the unknown phase $\phi$ reduces to identifying the control parameter $\theta$ that maximizes constructive interference. Within the CQL framework, this objective is encoded through the loss function $\mathcal{L}(\theta,\phi) = \sin^2\left(\frac{\theta-\phi}{2}\right),$ which attains its global minimum when $\theta=\phi$.

The parameter register representing the candidate values of $\theta$ is encoded using five qubits and initialized in a uniform superposition. The subsequent evolution is simulated coherently, following the CQL dynamics introduced in this work.  Details of the numerical implementation and simulation parameters are provided in Appendix~\ref{sec:interferometer_details}.

Fig.~\ref{fig:interferometer_results} illustrates the coherent evolution of the CQL parameter register throughout the optimization process.  As the evolution progresses, the increasing influence of the loss potential and the simultaneous suppression of kinetic spreading drive the wavefunction toward the global minimum of the interferometric loss landscape. By the end of the evolution, the probability amplitude is strongly localized around the optimal parameter value, corresponding to the unknown phase shift $\phi$. This demonstrates that the coherent dynamics successfully perform phase estimation without requiring repeated classical optimization steps.

\textit{Conclusion and outlook.---} We have introduced Coherent Quantum Learning (CQL), a framework in which the training of quantum models is itself implemented as a coherent quantum process. Model parameters are promoted to quantum degrees of freedom and evolved under a Hamiltonian encoding the objective function, replacing iterative classical optimization with unitary dynamics in parameter space. In this formulation, superposition and interference become active computational resources that shape the learning dynamics, allowing parameter configurations to be explored collectively rather than sequentially. More broadly, our perspective aligns with the long-standing view that information and physical dynamics are fundamentally intertwined~\cite{wheeler1990information,landauer1991information}. Here we suggest that learning itself may likewise be regarded as a physical process, realized through coherent quantum evolution.

Our construction provides a general and architecture-independent framework for quantum learning. The protocol applies to arbitrary parameterized quantum circuits, naturally accommodates batched training strategies, and is compatible with fault-tolerant quantum computing architectures where long coherent evolutions can be sustained. Numerical simulations demonstrate that the resulting dynamics progressively concentrate probability amplitude around low-loss regions of the landscape, effectively encoding learned representations within the quantum state itself. Several directions remain open for future investigation, including complexity analyses, improved Hamiltonian constructions and discretization strategies, and integration with error mitigation techniques~\cite{cai2023quantumerrormitigation}.

Overall, this work points toward a reconsideration of how learning should be formulated in quantum systems. Much of the current effort in quantum machine learning seeks to reproduce classical training paradigms, such as gradient-based optimization, within quantum settings. However, these paradigms implicitly impose a classical structure on the learning process, potentially obscuring the role of coherence and interference. In contrast, the protocol introduced here treats learning itself as a physical process governed by quantum dynamics, suggesting that quantum models may benefit from training principles that are native to quantum mechanics rather than adapted from classical algorithms.\\

\textit{Code and data availability}: The code and data for the numerical results are available at \url{https://github.com/nacedob/Coherent-Quantum-Learning}
\vspace{0.2cm}

\textit{Statement of AI usage}: We used basic AI tools for polishing code, enhancing figure design, and improving the clarity and readability of the writing. All scientific content, analysis, results, and conclusions were developed and verified entirely by the authors.

\vspace{0.2cm}
{\em Acknowledgments:} We thank Xabier Gutierrez, Richard Kueng and José Ignacio Latorre for discussions. LP is supported by the National Research Foundation, Singapore through the National Quantum Office, hosted in A*STAR, under its Centre for Quantum Technologies Funding Initiative (S24Q2d0009). LP is also partially supported by A*STAR under its Young Investigator Research Grant (YIRG) M25N8c0131. PRG and JGC acknowledge support from HORIZON-CL4-2022-QUANTUM01-SGA project 101113946 OpenSuperQ-Plus100 of the EU Flagship on Quantum Technologies, the Spanish Ram\'on y Cajal Grant RYC-2020-030503-I, and the “Generaci\'on de Conocimiento” project Grant No. PID2021-125823NA-I00 funded by MICIU/AEI/10.13039/501100011033, by “ERDF Invest in your Future” and by FEDER EU. PRG acknowledges support from UPV/EHU Ph.D. Grant No. PIFG 22/25. We also acknowledge support from the Basque Government through Grants No. IT1470-22, the Elkartek project KUBIBIT - kuantikaren berrikuntzarako ibilbide teknologikoak (ELKARTEK25/79), and from the IKUR Strategy under the collaboration agreement between Ikerbasque Foundation and BCAM on behalf of the Department of Education of the Basque Government. This work has also been partially supported by the Ministry for Digital Transformation and the Civil Service of the Spanish Government through the QUANTUM ENIA project call – Quantum Spain project, and by the European Union through the Recovery, Transformation and Resilience Plan – NextGenerationEU within the framework of the Digital Spain 2026 Agenda. We also acknowledge the European Quantum AI Summer School, where this project was initiated, and the CQT Hackamonth, where it was further developed.

\bibliography{bibliography}

\clearpage
\onecolumngrid

\begin{center}
  {\Large \textbf{Supplemental Material}}\\[0.25cm]
  {\normalsize for}\\[0.2cm]
  {\large \textbf{Quantum Hamiltonian Evolution for Coherent Quantum Learning}}\\[0.35cm]

  {\normalsize
Ignacio B. Acedo$^{1,*}$\\
Javier Gonzalez-Conde$^{1,2,3}$\\
Pablo Rodriguez-Grasa$^{2,4}$\\
Barry C. Sanders$^{5}$\\
and Lirandë Pira$^{6,\dagger}$}\\[0.25cm]

{\footnotesize\itshape
$^{1}$Quantum Mads, Calle Larrauri 1, Edificio A, piso 3, puerta 28, 48160 Derio, Spain\\
$^{2}$Department of Physical Chemistry, University of the Basque Country UPV/EHU, Apartado 644, 48080 Bilbao, Spain\\
$^{3}$EHU Quantum Center, University of the Basque Country UPV/EHU, Apartado 644, 48080 Bilbao, Spain\\
$^{4}$TECNALIA, Basque Research and Technology Alliance (BRTA), 48160 Derio, Spain\\
$^{5}$Institute for Quantum Science and Technology, University of Calgary, Alberta T3A 0E1, Canada\\
$^{6}$Centre for Quantum Technologies, National University of Singapore, Singapore\\[0.15cm]
$^{*}$Email: \texttt{ibenito@quantum-mads.com}\\
$^{\dagger}$Email: \texttt{lpira@nus.edu.sg}}
\end{center}

\vspace{0.5cm}

\noindent In this supplemental material, we provide technical details supporting the main text. We first describe the construction of the Hamiltonian underlying Coherent Quantum Learning (CQL) in \cref{sec:hamiltonian_construction}, including the implementation of functions of the position operator and the momentum operator. We then detail the construction of the cost Hamiltonian in \cref{sec:cost_hamiltonian}, presenting its extension to multiple simultaneously evaluated data points, the block-encoding of different cost functions, and the homogenization of block encodings required for Hamiltonian simulation. Next, we describe the batched CQL algorithm in \cref{sec:batched_cql}. Finally, we provide additional details on the numerical demonstrations presented in the main text: the quantum classification benchmark is discussed in \cref{sec:classification_details}, while the interferometric phase learning example, including the Mach--Zehnder interferometer evolution and simulation details, is presented in \cref{sec:interferometer_details}.

\section*{Contents}
\vspace{0.3cm}

\begin{enumerate}[S1.]
    \item \textit{Hamiltonian construction}
    \dotfill \pageref{sec:hamiltonian_construction}
    \begin{enumerate}[A.]
        \item \textit{Functions of the position operator}
        \dotfill \pageref{subsec:position_operator}
        \item \textit{Momentum operator}
        \dotfill \pageref{subsec:momentum_operator}
          \item \textit{Cost Hamiltonian}
        \dotfill \pageref{sec:cost_hamiltonian}
             \begin{enumerate}[1.]
             \item \textit{Extension to several simultaneously evaluated data points}
             \dotfill \pageref{subsec:multi_data}
             \item \textit{Block-encoding different cost functions}
             \dotfill \pageref{subsec:different_costs}
             \end{enumerate}
        \item \textit{Homogenizing block encodings}
        \dotfill \pageref{subsec:homogenizing}
    \end{enumerate}

    \item \textit{Batched CQL algorithm}
    \dotfill \pageref{sec:batched_cql}

    \item \textit{Details on quantum classification with a data re-uploading model}
    \dotfill \pageref{sec:classification_details}

    \item \textit{Details on interferometric phase learning}
    \dotfill \pageref{sec:interferometer_details}
    \begin{enumerate}[A.]
        \item \textit{Mach--Zehnder interferometer state evolution}
        \dotfill \pageref{subsec:mzi}
        \item \textit{Numerical implementation and simulation framework}
        \dotfill \pageref{subsec:numerics}
    \end{enumerate}
\end{enumerate}

\clearpage
\clearpage

\section{Hamiltonian Construction}\label{sec:hamiltonian_construction}
Our CQL Hamiltonian is time dependent. To overcome the explicit time dependence of the CQL Hamiltonian, we use the clock-dimension construction of Refs.~\cite{cao2023quantum,cao2024unifying}. The basic idea is to replace a non-autonomous Schrödinger equation by an autonomous one in a larger Hilbert space. Let the original parameter-space dynamics be generated by a time-dependent Hamiltonian, see Eq. (\ref{eq:ham_CQL}). The corresponding evolution is formally given by the time-ordered exponential
\begin{equation}
    U(T,0)
    =
    \mathcal T
    \exp\!\left(
        -i\int_0^T \hat H_{\mathrm{CQL}}(t)\,dt
    \right),
\end{equation}
which is difficult to implement directly because the Hamiltonian changes during the evolution. The clock construction removes this explicit time dependence by introducing an additional Hilbert space $\mathcal H_t$, with conjugate operators $\hat X_t$ and $\hat P_t$. The operator $\hat X_t$ stores the clock coordinate, while $\hat P_t=-i\partial_{x_t}$ generates translations along this coordinate. In the enlarged Hilbert space $ \mathcal H_t\otimes\mathcal H_\theta,$ the time-dependent scalar functions $e^{\varphi(t)}$ and $e^{\chi(t)}$ are promoted to diagonal operators $e^{\varphi(\hat X_t)}$ and $e^{\chi(\hat X_t)}$. This gives the time-independent Hamiltonian given by Eq. (\ref{eq:Hamil_time_indep}). The role of the first term, $\hat P_t\otimes\mathbb I$, is to translate the clock wavepacket through the clock register. If the clock register is initialized in a localized state $\ket{t_0}$, concentrated near $x_t=0$, then the translation generated by $\hat P_t$ makes the system sample the values of $e^{\varphi(x_t)}$ and $e^{\chi(x_t)}$ along the clock coordinate during the evolution. In this way, the original time dependence is encoded as position dependence in an auxiliary register. After discretizing the clock with $n_t$ qubits, the continuous operators $\hat X_t$ and $\hat P_t$ are replaced by finite-dimensional approximations acting on a grid of $2^{n_t}$ clock points.

This construction transfers the problem of implementing a time-dependent evolution to the problem of simulating the time-independent Hamiltonian in Eq.~\eqref{eq:Hamil_time_indep}. The price paid is the enlargement of the Hilbert space and the need to block-encode the new operators appearing in the clock Hamiltonian. In particular, to construct the full Hamiltonian one needs block-encodings of $\hat P_t$, $\hat P_\theta^2$, the diagonal functions $e^{\varphi(\hat X_t)}$ and $e^{\chi(\hat X_t)}$, and the loss Hamiltonian $\hat H_{\mathcal L}(x,\hat\theta)$. Once these block-encodings are available, the full Hamiltonian can be assembled using standard block-encoding arithmetic: tensor products for terms acting on different registers, multiplication for operator products, and linear-combination-of-unitaries techniques for the final sum.

\subsection{Functions of the Position Operator}\label{subsec:position_operator}
For our purposes, we are interested in block-encodings of diagonal functions of the position operator, $\hat X$, such as $e^{\varphi(\hat X)} \text{ and }e^{\chi(\hat X)}$ . Since $\hat X$ is diagonal on a uniform grid, any function $f(\hat X)$ is also diagonal:
\begin{equation}
    f(\hat X)
    =
    \operatorname{diag}\bigl(f(x_0),f(x_1),\ldots,f(x_{N-1})\bigr).
\end{equation}
One systematic way to construct such a block-encoding is to start from a block-encoding of the normalized position operator $\hat X/\alpha_X$, with spectrum contained in $[-1,1]$, and then use QSVT to implement a polynomial approximation $P_d$ to the desired function. More precisely, after mapping the spatial interval $[x_{\min},x_{\max}]$ to $[-1,1]$, one chooses a polynomial $P_d$ such that
\begin{equation}
    \max_{x\in[x_{\min},x_{\max}]}
    \left|
        P_d(x)-\frac{f(x)}{\alpha_f}
    \right|
    \leq \varepsilon ,
    \qquad
    f(x)=e^{\varphi(x)}
    \ \text{or}\
    f(x)=e^{\chi(x)},
\end{equation}
where $  \alpha_f \geq \max_{x\in[x_{\min},x_{\max}]} |f(x)|$ is the block-encoding normalization. Applying QSVT to the block-encoding of $\hat X/\alpha_X$ then gives a block-encoding of $  P_d(\hat X)\approx \frac{f(\hat X)}{\alpha_f}.$ The query complexity is linear in the polynomial degree $N_{\mathrm{queries}}=O(d)$ uses of the block-encoding of $\hat X/\alpha_X$ and its inverse, together with $O(d)$ single-qubit phase rotations, plus the gate cost of implementing the underlying block-encoding of $\hat X$. Therefore, the efficiency of this route is controlled by the approximation degree $d$. If $f$ is analytic on and around the relevant interval, Chebyshev or minimax polynomial approximations can achieve $ d = O(\log(1/\varepsilon))$ up to constants depending on the size of the analytic continuation region and the variation of $f$. By contrast, for functions with only finite smoothness, the degree may scale polynomially in $1/\varepsilon$. Thus, QSVT gives an asymptotically efficient block-encoding when the functions $e^{\varphi(x)}$ and $e^{\chi(x)}$ are sufficiently regular, but the cost can increase significantly for rapidly varying or nonsmooth coefficients.

There are also alternatives that exploit the fact that $f(\hat X)$ is diagonal. One possibility is to construct a direct LCU block-encoding using a Walsh--Hadamard or Pauli-$Z$ expansion of the sampled function on the grid:
\begin{equation}
    f(\hat X)
    =
    \sum_{S\subseteq\{1,\ldots,n\}}
    c_S Z_S,
    \qquad
    Z_S=\bigotimes_{j=1}^n Z_j^{\mathbf{1}_{j\in S}} .
\end{equation}
For the linear position operator, only $O(n)$ such Pauli-$Z$ strings are needed, which leads to a very efficient LCU construction as in Ref.~\cite{guseynov2025gateconstructionofblockencodings}. For a general function $f(x)$, however, the Walsh expansion may contain many nonzero coefficients, potentially up to $2^n$, unless $f$ has additional structure or the expansion can be truncated efficiently. In that case, the LCU cost is controlled by the number of retained Walsh coefficients and by their
$\ell_1$-norm,
\begin{equation}
    \alpha_{\mathrm{LCU}}=\sum_S |c_S|.
\end{equation}

Another direct approach is to compute the grid value $f(x_j)$ reversibly and load it into an ancilla amplitude. Namely, one implements an arithmetic oracle of the form
\begin{equation}
    |j\rangle|0\rangle
    \longmapsto
    |j\rangle
    \left(
        \frac{f(x_j)}{\alpha_f}|0\rangle
        +
        \sqrt{1-\left|\frac{f(x_j)}{\alpha_f}\right|^2}|1\rangle
    \right),
\end{equation}
which directly block-encodes the diagonal matrix $f(\hat X)/\alpha_f$. The cost of this construction is governed by the reversible arithmetic required to evaluate $f(x_j)$ to the desired precision, rather than by a QSVT polynomial degree. This can be preferable when $f$ has an efficient arithmetic description, for example when $\varphi(x)$ or $\chi(x)$ are simple polynomials, rational functions, or elementary functions. Therefore, QSVT is a natural and general route when one already has a good block-encoding of $\hat X$ and a low-degree polynomial approximation to $f$, whereas direct Walsh/LCU or arithmetic amplitude-loading constructions may be more efficient for diagonal functions with exploitable structure.

\subsection{Momentum Operator}\label{subsec:momentum_operator}

The choice for representing the momentum operator $\hat P$ on a discrete space is not unique and depends on the numerical scheme employed and the boundary conditions of interest, each leading to a different matrix representation with distinct accuracy, sparsity, and stability properties.  For instance, finite-difference discretizations produce sparse matrices whose entries depend explicitly on the boundary treatment, whereas Fourier spectral methods produce a diagonal representation in momentum space that is generally dense in position space. Consequently, a block encoding of the momentum operator should always be understood as a block encoding of a specific discretized version of the continuous operator, rather than of a unique finite-dimensional object.

A common drawback of these discretizations is the scaling of the norm of the momentum operator. In the absence of additional regularity assumptions, the derivative operator is unbounded. Consequently, refining the grid increases the largest momentum mode that can be represented and, therefore, increases the normalization factor required in a block-encoding. This directly affects the cost of Hamiltonian simulation. In typical discretizations, the norm of the discretized momentum operator scales linearly with the number of grid points, that is, as $O(N)$. Since $N=2^n$ for an $n$-qubit spatial register, this corresponds to an exponential scaling in the number of qubits. An explicit methodology for achieving a block encoding of the momentum operator can be found in Ref. \cite{guseynov2025gateconstructionofblockencodings}.

However, the unfavorable scaling is not intrinsic to all implementations, but rather to treating the derivative operator as a generic matrix whose block-encoding normalization is set by its worst-case operator norm. Since the derivative is unbounded, this worst-case norm necessarily grows as the grid is refined. To avoid paying this cost, one must either exploit additional structure of the derivative operator or impose regularity assumptions on the states being simulated.

To mitigate this scaling, we need to explore specific structures of the derivative matrices. The idea relies in the fact that the large norm of the discretized derivative is caused by high-frequency grid modes, restricting the dynamics to a subspace $|p|\leq K$ replaces the worst-case bound $ \|P\|=O(N)$ by an effective bound $    \|P\|_{\mathrm{eff}}=O(K).$ For sufficiently smooth or analytic solutions, $K$ can be much smaller than the total number of grid points $N$, so the Hamiltonian-simulation cost is controlled by the physically occupied bandwidth rather than by the formal ultraviolet cutoff of the discretization. \\

In order to illustrate how to achieve a block encoding of this operator (without mitigating the scaling), we can consider a periodic grid. Let $S$ denote the cyclic shift
\begin{equation}
    S|j\rangle=|j+1 \!\!\!\pmod N\rangle.
\end{equation}
The central-difference momentum operator can be written as
\begin{equation}
    \hat P
    =
    -i\frac{S-S^\dagger}{2\Delta x}
    =
    \frac{1}{2\Delta x}
    \left(
        -iS+iS^\dagger
    \right).    
\end{equation}
Since $-iS$ and $iS^\dagger$ are unitary, this gives an immediate LCU block-encoding. With $\mathrm{PREP}|0\rangle=\frac{|0\rangle+|1\rangle}{\sqrt 2}$ and $\mathrm{SELECT}=|0\rangle\langle 0|\otimes (-iS)+|1\rangle\langle 1|\otimes (iS^\dagger),$ one obtains
\begin{equation}
    (\langle +|\otimes I)\mathrm{SELECT}(|+\rangle\otimes I)
    =
    \Delta x\,\hat P .   
\end{equation}
Thus this construction gives a $\left(\frac{1}{\Delta x},1,0\right)$ block-encoding of $\hat P$. Note that the rescaling factor $\frac{1}{\Delta x}$ will lead to a simulation time scaling like $t\sim \Delta x$ (exponential with number of qubits).

A block-encoding of $\hat P^2$ can be obtained in two natural ways. One may either compose two block-encodings of $\hat P$, using the standard block-encoding multiplication rule, or construct a block-encoding of $\hat P^2$ directly by exploiting the explicit finite-difference stencil of the second-derivative operator.

\subsection{Cost Hamiltonian}\label{sec:cost_hamiltonian}
We now describe the construction of the loss Hamiltonian. The objective is to block-encode
\begin{equation}
    \hat H_{\mathcal L}(x,\hat\theta)
    =
    \sum_{\theta\in\mathcal G_m}
    \mathcal L(x,\theta)
    \ket{\theta}\bra{\theta},
    \label{eq:loss_hamiltonian_objective}
\end{equation}
which is diagonal in the parameter register. The key primitive is the coherent model-evaluation unitary $U_{x,\theta}$, which evaluates the quantum model simultaneously over all grid values of the trainable parameter. Acting on the model qubit and the $m$-qubit parameter register, it prepares
\begin{equation}
    U_{x,\theta}
    \bigl(\ket{0}\otimes\ket{0}^{\otimes m}\bigr)
    =
    \sum_{\theta\in\mathcal G_m}
    \left[
        \alpha_0(x,\theta)\ket{0}\ket{\theta}
        +
        \alpha_1(x,\theta)\ket{1}\ket{\theta}
    \right].
    \label{eq:coherent_model_evaluation}
\end{equation}
Thus the amplitudes of the model output are coherently correlated with the parameter values $\theta$. For binary classification, and without loss of generality taking $y=1$, the loss is
\begin{equation}
    \mathcal L(x,\theta)
    =
    1-|\alpha_1(x,\theta)|^2
    =
    |\alpha_0(x,\theta)|^2.
    \label{eq:binary_classification_loss}
\end{equation}
Therefore, it is enough to construct a diagonal operator whose entries are $|\alpha_0(x,\theta)|^2$.

\begin{figure}[t!]
    \centering
    \includegraphics[width=.3\textwidth]{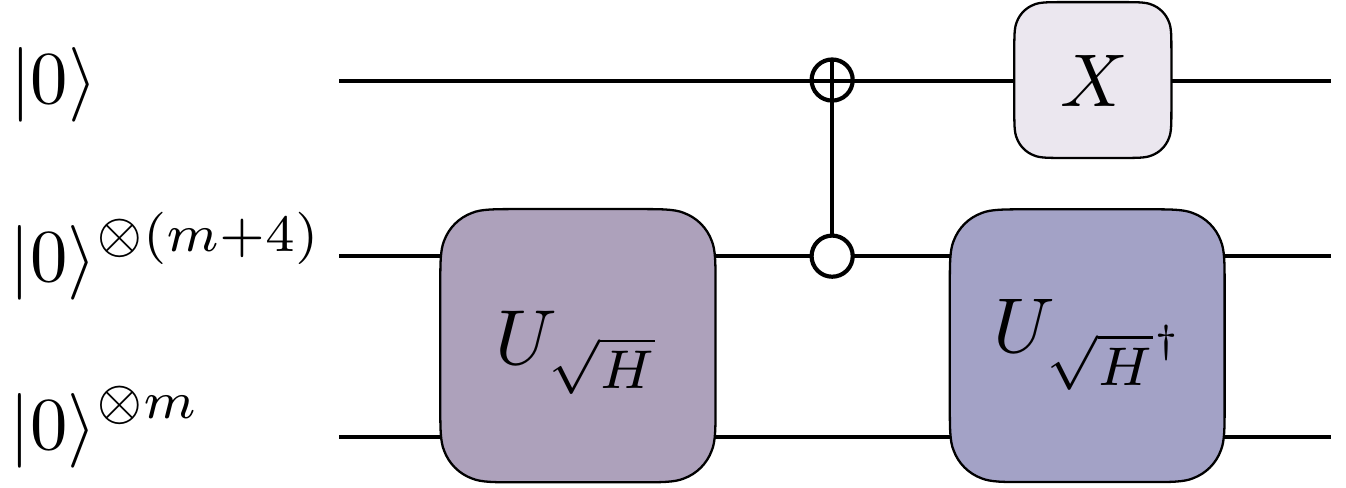}
    \caption{Circuit that implements the block encoding of $H$ from the block encodings of $\sqrt{H}$ and $\sqrt{H}^\dagger$. It requires access to the controlled block encoding of the roots of $H$, one more ancilla, an (m+4)-controlled generalized Toffoli gate and one $X$ gate.
    }
    \label{fig:appendix_multiplication_BE}
\end{figure}

To do this, we first construct a block-encoding of the ``square-root'' diagonal operator
\begin{equation}
    \sqrt H
    =
    \sum_{\theta\in\mathcal G_m}
    \alpha_0(x,\theta)
    \ket{\theta}\bra{\theta}.
    \label{eq:sqrt_loss_operator}
\end{equation}
Using the amplitude-to-diagonal conversion of Ref.~\cite{rattew2023nonlinear}, query access to a controlled version of $U_{x,\theta}$ gives a $(1,m+4,0)$ block-encoding $U_{\sqrt H}$ of $\sqrt H$, with circuit depth $\mathcal O(m)$ and $\mathcal O(1)$ queries to the controlled model evaluation unitary. The ancilla used to store the classification outcome is included in the ancilla count. Applying the same construction to $U_{x,\theta}^\dagger$ gives a block-encoding $U_{\sqrt H^\dagger}$ of
\begin{equation}
    \sqrt H^\dagger
    =
    \sum_{\theta\in\mathcal G_m}
    \alpha_0^*(x,\theta)
    \ket{\theta}\bra{\theta}.
    \label{eq:sqrt_loss_adjoint_operator}
\end{equation}
Since the loss Hamiltonian is
\begin{equation}
    \hat H_{\mathcal L}(x,\hat\theta)
    =
    \sqrt H\,\sqrt H^\dagger
    =
    \sum_{\theta\in\mathcal G_m}
    |\alpha_0(x,\theta)|^2
    \ket{\theta}\bra{\theta},
    \label{eq:loss_hamiltonian_factorization}
\end{equation}
we obtain a block-encoding of $\hat H_{\mathcal L}$ by multiplying the two block-encodings, see Fig.~\ref{fig:appendix_multiplication_BE}. Using the standard block-encoding product rule, with the two encodings acting on compatible ancilla spaces, this multiplication requires one additional ancilla qubit. Therefore, the resulting block-encoding of the loss Hamiltonian has parameters $(1,m+5,0)$ with overall circuit depth $\mathcal O(m)$ and $\mathcal O(1)$ queries to a controlled version of $U_{x,\theta}$.

\subsubsection{Extension to several simultaneously evaluated data points}\label{subsec:multi_data}
The construction above can be extended to a batch of data points evaluated coherently in the same circuit. For clarity, consider first a batch of two samples
\begin{equation}
    \mathcal B=\{(x_1,y_1),(x_2,y_2)\}.
    \label{eq:two_sample_batch}
\end{equation}
Instead of using a single model-output qubit, we introduce two output qubits, one for each data point. The coherent model-evaluation unitary now acts as
\begin{equation}
    U_{\mathcal B,\theta}
    \left(
        |0\rangle|0\rangle \otimes |0\rangle^{\otimes m}
    \right)
    =
    \sum_{\theta\in\mathcal G_m}
    |\alpha(x_1,\theta)\rangle
    \otimes
    |\alpha(x_2,\theta)\rangle
    \otimes
    |\theta\rangle ,
    \label{eq:batch_model_evaluation_state}
\end{equation}
where
\begin{equation}
    |\alpha(x_b,\theta)\rangle
    =
    \alpha_0(x_b,\theta)|0\rangle
    +
    \alpha_1(x_b,\theta)|1\rangle ,
    \qquad b=1,2.
    \label{eq:single_sample_output_state_batch}
\end{equation}
Equivalently,
\begin{equation}
    U_{\mathcal B,\theta}
    \left(
        |00\rangle \otimes |0\rangle^{\otimes m}
    \right)
    =
    \sum_{\theta\in\mathcal G_m}
    \sum_{z_1,z_2\in\{0,1\}}
    \alpha_{z_1}(x_1,\theta)
    \alpha_{z_2}(x_2,\theta)
    |z_1z_2\rangle|\theta\rangle .
    \label{eq:batch_model_evaluation_expanded}
\end{equation}

If the labels are $y_1,y_2$, the amplitude associated with classifying both examples correctly is
\begin{equation}
    \beta_{\mathcal B}(\theta)
    =
    \alpha_{y_1}(x_1,\theta)
    \alpha_{y_2}(x_2,\theta).
    \label{eq:batch_correct_classification_amplitude}
\end{equation}
By applying $X$ gates to those output qubits whose label is $1$, the correct label string $|y_1y_2\rangle$ can be mapped to $|00\rangle$. Thus, after this relabeling, the amplitude of the state $|00\rangle$ is precisely $\beta_{\mathcal B}(\theta)$.

Using the same amplitude-to-diagonal conversion as in the single-sample case, query access to $U_{\mathcal B,\theta}$ gives a block-encoding of the diagonal operator
\begin{equation}
    \sqrt H_{\mathcal B}
    =
    \sum_{\theta\in\mathcal G_m}
    \beta_{\mathcal B}(\theta)
    |\theta\rangle\langle\theta| .
    \label{eq:batch_sqrt_operator}
\end{equation}
Multiplying this block-encoding by its adjoint gives
\begin{equation}
    H_{\mathcal B}^{\mathrm{corr}}
    =
    \sqrt H_{\mathcal B}\sqrt H_{\mathcal B}^{\dagger}
    =
    \sum_{\theta\in\mathcal G_m}
    |\beta_{\mathcal B}(\theta)|^2
    |\theta\rangle\langle\theta| .
    \label{eq:batch_correct_classification_hamiltonian}
\end{equation}
Since
\begin{equation}
    |\beta_{\mathcal B}(\theta)|^2
    =
    |\alpha_{y_1}(x_1,\theta)|^2
    |\alpha_{y_2}(x_2,\theta)|^2 ,
    \label{eq:batch_correct_probability_factorization}
\end{equation}
this operator encodes the probability that both samples in the batch are classified correctly:
\begin{equation}
    H_{\mathcal B}^{\mathrm{corr}}
    =
    \sum_{\theta\in\mathcal G_m}
    p_{y_1}(x_1,\theta)
    p_{y_2}(x_2,\theta)
    |\theta\rangle\langle\theta| .
    \label{eq:batch_correct_probability_operator}
\end{equation}
Therefore, a natural joint batch loss is
\begin{equation}
    \hat H_{\mathcal L}^{\mathrm{joint}}
    =
    I-H_{\mathcal B}^{\mathrm{corr}}
    =
    \sum_{\theta\in\mathcal G_m}
    \left[
        1-
        p_{y_1}(x_1,\theta)p_{y_2}(x_2,\theta)
    \right]
    |\theta\rangle\langle\theta| .
    \label{eq:joint_batch_loss_hamiltonian}
\end{equation}
This construction is a direct generalization of the single-sample loss. However, it should be noted that it corresponds to a product-type batch objective, not to the usual empirical mean loss.

\subsubsection{Block-encoding different cost functions}\label{subsec:different_costs}
The construction of the loss Hamiltonian can be generalized to a broad class of cost functions by observing that, in the CQL framework, the objective function always appears as a diagonal operator on the parameter register,
\begin{equation}
    \hat H_{\mathcal L}
    =
    \sum_{\theta\in\mathcal G_m}
    \mathcal L(\theta)
    |\theta\rangle\langle\theta| .
\end{equation}
Thus, constructing a block-encoding of a cost function amounts to constructing a block-encoding of a diagonal matrix whose entries are the loss values over the discretized parameter grid.

The basic primitive is the coherent model-evaluation unitary
\begin{equation}
    U_{x,\theta}
    \bigl(|0\rangle\otimes |0\rangle^{\otimes m}\bigr)
    =
    \sum_{\theta\in\mathcal G_m}
    \left[
        \alpha_0(x,\theta)|0\rangle|\theta\rangle
        +
        \alpha_1(x,\theta)|1\rangle|\theta\rangle
    \right],
\end{equation}
which evaluates the model simultaneously for all parameter values $\theta$. For a binary classification task with label $y$, the relevant probability is
\begin{equation}
    p_y(x,\theta)
    =
    |\alpha_y(x,\theta)|^2 .
\end{equation}
Therefore, many common losses can be written as a scalar function of this probability,
\begin{equation}
    \mathcal L_y(x,\theta)
    =
    F_y\!\left(p_y(x,\theta)\right).
\end{equation}
The block-encoding strategy then consists of two steps: first construct a block-encoding of the probability operator
\begin{equation}
    \hat P_y
    =
    \sum_{\theta\in\mathcal G_m}
    p_y(x,\theta)
    |\theta\rangle\langle\theta|,
\end{equation}
and then apply a polynomial or arithmetic transformation that maps $p_y\mapsto F_y(p_y)$.

For example, using amplitude-to-diagonal conversion, query access to $U_{x,\theta}$ can be used to construct a block-encoding of
\begin{equation}
    \sqrt P_y
    =
    \sum_{\theta\in\mathcal G_m}
    \alpha_y(x,\theta)
    |\theta\rangle\langle\theta|.
\end{equation}
Multiplying this block-encoding by its adjoint gives
\begin{equation}
    \sqrt P_y\,\sqrt P_y^\dagger
    =
    \sum_{\theta\in\mathcal G_m}
    |\alpha_y(x,\theta)|^2
    |\theta\rangle\langle\theta|
    =
    \hat P_y .
\end{equation}
Once $\hat P_y$ is block-encoded, different losses can be obtained by applying a function $F_y$ to this diagonal operator:
\begin{equation}
    \hat H_{\mathcal L}
    =
    F_y(\hat P_y)
    =
    \sum_{\theta\in\mathcal G_m}
    F_y(p_y(x,\theta))
    |\theta\rangle\langle\theta| .
\end{equation}

A first important family consists of polynomial losses. If $F_y$ is a polynomial of degree $d$,
\begin{equation}
    F_y(p)=\sum_{k=0}^d c_k p^k,
\end{equation}
then $F_y(\hat P_y)$ can be block-encoded using QSVT or by combining powers of the block-encoding of $\hat P_y$. In this case, the query complexity scales linearly with the polynomial degree:
\begin{equation}
    N_{\mathrm{queries}}
    =
    \mathcal O(d)
\end{equation}
uses of the block-encoding of $\hat P_y$ and its inverse, up to the cost of preparing the required QSVT phase sequence or LCU coefficients. This immediately covers losses such as
\begin{equation}
    \mathcal L_y(x,\theta)=1-p_y(x,\theta),
\end{equation}
which has degree one, and mean-square-type losses such as
\begin{equation}
    \mathcal L_y(x,\theta)=\left(1-p_y(x,\theta)\right)^2,
\end{equation}
which have degree two.

More general smooth losses can be handled by polynomial approximation. Suppose $F_y$ is bounded on the relevant interval $p\in[0,1]$, and choose a polynomial $P_d$ such that
\begin{equation}
    \max_{p\in[0,1]}
    \left|
        P_d(p)-\frac{F_y(p)}{\alpha_F}
    \right|
    \leq \varepsilon ,
\end{equation}
where
\begin{equation}
    \alpha_F\geq \max_{p\in[0,1]} |F_y(p)|
\end{equation}
is the block-encoding normalization. Applying QSVT to the block-encoding of $\hat P_y$ gives
\begin{equation}
    P_d(\hat P_y)
    \approx
    \frac{F_y(\hat P_y)}{\alpha_F}.
\end{equation}
The cost is
\begin{equation}
    N_{\mathrm{queries}}=\mathcal O(d).
\end{equation}
If $F_y$ is analytic on and around the interval of interest, the degree can often scale as
\begin{equation}
    d=\mathcal O(\log(1/\varepsilon)),
\end{equation}
up to constants depending on the analytic domain. If $F_y$ is non smooth or has singularities, the polynomial degree may scale polynomially in $1/\varepsilon$. For instance, a cross-entropy-type loss
\begin{equation}
    \mathcal L_y(x,\theta)=-\log p_y(x,\theta)
\end{equation}
requires restricting the domain to
\begin{equation}
    p_y(x,\theta)\geq \delta>0
\end{equation}
or regularizing it as
\begin{equation}
    \mathcal L_y(x,\theta)=-\log\!\left(p_y(x,\theta)+\delta\right),
\end{equation}
because the logarithm is singular at $p=0$. The approximation degree then depends not only on $\varepsilon$, but also on the cutoff parameter $\delta$.

Alternatively, the loss might be expressed as the expectation value of an observable $O$ on the model evaluation qubit register. Suppose
\begin{equation}
    \mathcal L(x,\theta)
    =
    \langle \psi(x,\theta)|O|\psi(x,\theta)\rangle ,
\end{equation}
where $O$ is a Hermitian observable with an efficient block-encoding or LCU decomposition. Since
\[
    |\psi(x,\theta)\rangle
\]
is prepared coherently by the model-evaluation unitary, one can use a Hadamard-test-like or amplitude-to-diagonal construction to encode the diagonal operator
\begin{equation}
    \hat H_{\mathcal L}
    =
    \sum_{\theta\in\mathcal G_m}
    \langle \psi(x,\theta)|O|\psi(x,\theta)\rangle
    |\theta\rangle\langle\theta| .
\end{equation}
If the final cost is a nonlinear function of this expectation value, the same QSVT or polynomial-approximation strategy can then be applied.

\subsection{Homogenizing Block Encodings}\label{subsec:homogenizing}
Suppose $U_A$ is an $(\alpha_A,a,\varepsilon_A)$-block-encoding of $A$ and $U_B$ is an $(\alpha_B,b,\varepsilon_B)$-block-encoding of $B$. Let $m=\max(a,b)$ and pad the smaller block-encoding with dummy ancillas so that both use $m$ ancilla qubits. Then choose $\alpha\geq \max\{\alpha_A,\alpha_B\}$. For $X\in\{A,B\}$, define $\gamma_X=\frac{\alpha_X}{\alpha}$. Adding one extra ancilla and a single-qubit unitary $R_X$ satisfying $\langle 0|R_X|0\rangle=\gamma_X$, the unitary $\overline U_X = R_X\otimes \widetilde U_X$ satisfies
\begin{equation}
    (\langle 0^{m+1}|\otimes I)\overline U_X
    (|0^{m+1}\rangle\otimes I)
    =
    \frac{X}{\alpha}.
\end{equation}
Thus, both block-encodings can be homogenized to have the same number of ancillas and the same normalization $\alpha$. Once homogenized we can sum the three terms using standard LCU.

\section{Batched CQL Algorithm}\label{sec:batched_cql}
This appendix provides the core physical and mathematical details for the batched Coherent Quantum Learning CQL framework. To maintain a direct and transparent presentation, we omit implementation-specific hardware details, such as the auxiliary clock register dimension and block-encoding circuits, focusing instead on the continuous state evolution which applies uniformly across all batches. The complete operational implementation of this framework is structured as follows:\\

\begin{algorithm}[H]
\caption{Batched Coherent Quantum Learning (CQL)}
\label{alg:simple_cql}
\LinesNotNumbered
\SetKwInOut{Input}{Input}
\SetKwInOut{Output}{Output}

\Input{
    Dataset $\mathcal{D}$ split into $B$ batches $\{\mathcal{B}_1, \mathcal{B}_2, \dots, \mathcal{B}_B\}$; \\
    Total training epochs $E$; \\
    Evolution time per batch $\tau$.
}
\Output{Optimized parameter state $\ket{\Psi_{\text{final}}}$.}

\BlankLine
\tcp{1. Setup Initial Search Space}
\If{no prior information available}{
    Initialize parameter state $\ket{\Psi}$ as a flat, uniform superposition across all possible parameter options $\theta$\;
}
\Else{
    Initialize parameter state $\ket{\Psi}$ as a localized Gaussian wavepacket centered at an informed initial guess $\theta_0$\;
};

\BlankLine
\tcp{2. Main Training Loops}
\For{\text{epoch } $e = 1$ \KwTo $E$}{
    \text{Shuffle dataset batches}\;
    
    \For{\text{batch } $b = 1$ \KwTo $B$}{
        Read current data samples from batch $\mathcal{B}_b$\;
        
        Map the batch's cost function into the physical system as a loss landscape operator $\hat{H}_{\mathcal{L}}(\mathcal{B}_b)$\;
        
        Combine Kinetic and Loss forces into the total batch Hamiltonian $\hat{H}_{\text{CQL}}^{(b)}(t)$\;
        
        \tcp{Let the physics naturally update the parameters}
        Evolve the quantum state continuously for duration $\tau$:\\
        $\ket{\Psi} \leftarrow \text{Evolve}\left( \ket{\Psi}, \hat{H}_{\text{CQL}}^{(b)}, \tau \right)$\;
        
        \BlankLine
        \textit{Note: The resulting state $\ket{\Psi}$ is passed directly as the starting point for the next batch (No classical measurement intermediary!)}\;
    }
}

\BlankLine
$\ket{\Psi_{\text{final}}} \leftarrow \ket{\Psi}$\;
\Return{$\ket{\Psi_{\text{final}}}$}
\end{algorithm}

The training dataset $\mathcal{D}$ is partitioned into $B$ mini-batches $\{\mathcal{B}_1, \dots, \mathcal{B}_B\}$. The continuous model parameters are mapped onto an $m$-qubit discrete coordinate grid $\mathcal{G}_m$ containing $2^m$ states. For each batch $\mathcal{B}_b$, we define a diagonal loss Hamiltonian representing the potential energy landscape of that specific batch:
\begin{equation}
    \hat{H}_{\mathcal{L}}(\mathcal{B}_b) = \sum_{\theta \in \mathcal{G}_m} \mathcal{L}(\mathcal{B}_b, \theta) \ket{\theta}\bra{\theta}
\end{equation}
where $\mathcal{L}(\mathcal{B}_b, \theta)$ is the batch loss evaluated at coordinate $\theta$.

The training process is driven by a time-dependent physical system combining kinetic exploration $\hat{K}$ and batch-specific potential attraction $\hat{H}_{\mathcal{L}}(\mathcal{B}_b)$. For any batch $b$, the system evolves under the Hamiltonian:
\begin{equation}
    \hat{H}_{\text{CQL}}^{(b)}(t) = e^{\varphi(t)} \hat{K} + e^{\chi(t)}\hat{H}_{\mathcal{L}}(\mathcal{B}_b)
\end{equation}

The quantum state vector is propagated sequentially from one batch to the next. The final state of batch $b-1$ serves directly as the initial state for batch $b$:
\begin{equation}
    \ket{\Psi_{b}} = \mathcal{T}\exp\left(-i \int_0^\tau \hat{H}_{\text{CQL}}^{(b)}(t) \, dt\right) \ket{\Psi_{b-1}}
\end{equation}
Because the evolution time $\tau$ per batch is finite and the damping ratio $e^{\varphi(t)}/e^{\chi(t)}$ does not collapse to zero within a single batch, the wavefunction retains a non-zero kinetic dispersion. This physical mechanism introduces an implicit regularization, analogous to a classical learning rate, preventing the state from freezing into a single batch's local minimum, thereby avoiding catastrophic forgetting across the global training sequence.

\section{Details on Quantum Classification with a Data Re-Uploading Model}\label{sec:classification_details}
\begin{figure}[t]
    \centering
    \begin{subfigure}[b]{0.99\linewidth}
        \centering
        \includegraphics[width=0.55\linewidth]{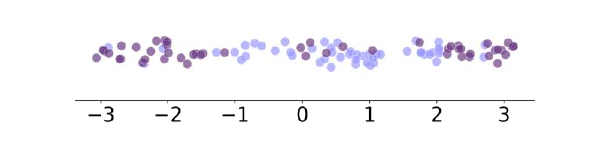}
        \caption{Dataset employed in the numerical simulations. The task is a one-dimensional binary classification problem in which samples lying inside the interval $\left(-\frac{\pi}{3}, \frac{2\pi}{3}\right)$ are assigned to one class, while the remaining samples belong to the other class. Noise is added to the data. The training and test sets contain 100 and 500 samples, respectively.}
        \label{fig:appendix_dataset}
    \end{subfigure}
    \vspace{0.5cm} 
    \begin{subfigure}[b]{0.95\linewidth}
        \centering
        \includegraphics[width=0.6\linewidth]{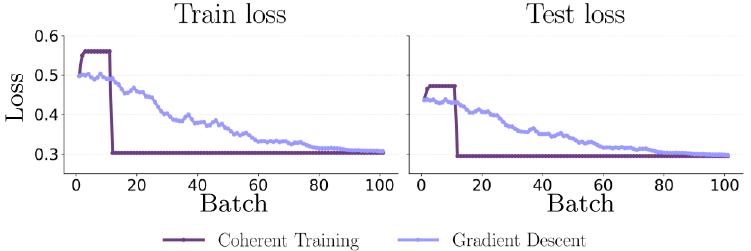}
        \caption{Comparison of the train and test loss evolution during training for a traditional hybrid gradient descent (lavender) and the proposed CQL (purple).}
        \label{fig:appendix_loss}
    \end{subfigure}

    \caption{Numerical simulations and training performance for the proposed CQL protocol.}
    \label{fig:appendix}
\end{figure}

In this section, we provide additional details on the numerical results presented in Fig.~\ref{fig:batch_illustration}. We consider a single-qubit data re-uploading model with two layers, resulting in two trainable parameters. The parameter space is discretized using five qubits per parameter, yielding a grid of $2^5$ points over the interval $(-\pi,\pi)$, corresponding to a resolution of $\frac{2\pi}{2^5} \simeq 0.20$ radians. The dataset, shown in Fig.~\ref{fig:appendix}(a), consists of a one-dimensional binary classification task in which the objective is to determine whether a data point lies inside the interval $\left(-\frac{\pi}{3}, \frac{2\pi}{3}\right)$ or outside it. To avoid trivial separability, additive noise is introduced by perturbing the input data, thereby broadening the class boundary. The training and test sets contain $100$ and $500$ samples, respectively; the larger test set is used to obtain more reliable performance estimates, as evaluation is computationally inexpensive compared to training.

The learning dynamics are governed by the damping functions
\begin{equation}\nonumber
\varphi(t)=\frac{1}{t^3},
\qquad
\chi(t)=t^3,
\end{equation}
which control the effective learning rate and convergence behavior. The total evolution time is set to $10^{-2},\mathrm{s}$, chosen heuristically to balance learning effectiveness and robustness to outliers. The initial state of $\hat{\theta}$ is taken as a uniform superposition over all discretized parameter values, reflecting the absence of prior information. For comparison, the same model is also trained using a classical optimizer, namely stochastic gradient descent with learning rate $0.1$ and random parameter initialization.

Fig.~\ref{fig:appendix}(b) shows the evolution of both training and test losses during sequential single-sample updates induced by the CQL protocol. A conventional hybrid optimization scheme based on stochastic gradient descent, using the same batch size of one, is included for comparison. The loss rapidly converges—within approximately $15$ updates—to a regime where the dominant minima are correctly identified, and the most probable sampled parameters correspond to near-optimal values. Beyond this point, further improvements in the loss are marginal. In contrast, the probability distribution over parameters exhibits a richer dynamical behavior, continuing to evolve after loss convergence. In particular, it progressively refines its structure and captures finer features of the landscape, thereby providing a more informative representation of the learning process than the loss alone.

\section{Details on Interferometric Phase Learning}\label{sec:interferometer_details}

\subsection{Mach–Zehnder Interferometer State Evolution}\label{subsec:mzi}
To rigorously model the evolution of a single photon injected into the horizontal input port from the left, we map the lower and upper spatial paths onto the standard single-qubit state space $\mathbb{C}^2$ where $|0\rangle = \begin{pmatrix} 1 & 0 \end{pmatrix}^T$ and $|1\rangle = \begin{pmatrix} 0 & 1 \end{pmatrix}^T$. A symmetric 50:50 beam splitter (BS) is characterized by the unitary operation:$$ BS = \frac{1}{\sqrt{2}} \begin{pmatrix} 1 & i \\ i & 1 \end{pmatrix},$$where the off-diagonal imaginary component $i$ captures the relative $90^\circ$ phase shift acquired upon reflection. The dual-path phase shifting stage is represented by the diagonal unitary operation $P = \text{diag}(e^{i\theta}, e^{i\phi})$.Beginning in the initial lower-path state $|\psi_i\rangle = |0\rangle$, the complete unitary evolution through the Mach–Zehnder interferometer sequence yields the final output state vector before measurement:\begin{align}|\psi_f\rangle &= BS \;P\; BS |0\rangle \nonumber \&= \frac{1}{2} \begin{pmatrix} 1 & i \\ i & 1 \end{pmatrix} \begin{pmatrix} e^{i\theta} & 0 \\ 0 & e^{i\phi} \end{pmatrix} \begin{pmatrix} 1 \\ i \end{pmatrix} \nonumber \&= \frac{1}{2} \begin{pmatrix} e^{i\theta} - e^{i\phi} \\ i(e^{i\theta} + e^{i\phi}) \end{pmatrix}.\end{align}By factoring out the common phase factor $i e^{i\frac{\phi + \theta}{2}}$ and exploiting Euler's trigonometric identities, the state vector can be compactly rewritten as a superposition of the computational basis kets:$$|\psi_f\rangle = i e^{i\frac{\phi + \theta}{2}} \left[ -\sin\left(\frac{\phi - \theta}{2}\right)|0\rangle + \cos\left(\frac{\phi - \theta}{2}\right)|1\rangle \right].$$The projection probabilities $p(0) = |\langle 0 | \psi_f \rangle|^2$ and $p(1) = |\langle 1 | \psi_f \rangle|^2$ follow immediately via Born's rule, yielding the standard interference formulas utilized in the main text.

\subsection{Numerical Implementation and Simulation Framework}\label{subsec:numerics}
The numerical evaluation of the path-encoded phase estimation framework under CQL is implemented by mapping the continuous parameter space $\theta \in [-\pi, \pi]$ onto a discrete computational grid. For an $n$-qubit representation ($n = 5$), the coordinate space is uniformly partitioned into $2^n = 32$ discrete points, while the temporal domain over a total duration $T = 5.0$ is sampled via $N = 1000$ discrete time slices, defining a temporal resolution of $\delta t = 5 \times 10^{-3}$. The target state vector corresponding to the unknown phase shift $\phi$ is initialized within the simulator using the equivalent two-level projection $|\psi_{\text{target}}\rangle = \begin{pmatrix} \cos(\phi/2) & -i\sin(\phi/2) \end{pmatrix}^T$, which governs the physical loss landscape $\mathcal{L}(\theta) = 1 - |\langle \psi_{\text{target}} | \psi(\theta) \rangle|^2$.  For the time-dependent schedules, we employ $\phi(t) = 1/t$ for the kinetic term and $\chi(t) = t$ for the potential contribution, which satisfy the requirements of the CQL protocols.

\end{document}